\documentclass[showpacs,pra,reprint,reprintnumbers,amsmath,amssymb,superscriptaddress]{revtex4-1}
\usepackage{graphicx}
\usepackage{dcolumn}
\usepackage{colortbl}
\usepackage{enumitem}

\begin{document}

\title{Nuclear Spin Squeezing via Electric Quadrupole Interaction}

\author{Ya\u{g}mur  \surname{Aksu Korkmaz}}
\author{Ceyhun \surname{Bulutay}}
\email{bulutay@fen.bilkent.edu.tr}
\affiliation{Department of Physics, Bilkent University, Ankara 06800, Turkey}

\date{\today}

\begin{abstract}
Control over nuclear spin fluctuations is essential for processes that rely on
preserving the quantum state of an embedded system. For this purpose, squeezing 
is a viable alternative, so far that has not been properly exploited for the nuclear spins. 
Of particular relevance in solids is the electric quadrupole interaction (QI), which operates 
on nuclei having spin higher than 1/2. In its general form, QI involves an electric field gradient 
(EFG) biaxiality term. Here, we show that as this EFG biaxiality increases, it enables continuous 
tuning of single-particle squeezing from the one-axis twisting to the two-axis countertwisting limits. 
A detailed analysis of QI squeezing is provided, exhibiting the intricate consequences of EFG 
biaxiality. The  initial states over the Bloch sphere are mapped out to identify those favorable for
fast initial squeezing, or for prolonged squeezings.
Furthermore, the evolution of squeezing in the presence of a phase-damping channel and an external
magnetic field are investigated. We observe that dephasing drives toward an anti-squeezed terminal 
state, the degree of which increases with the spin angular momentum. Finally, QI squeezing in the 
limiting case of a two-dimensional EFG with a perpendicular magnetic field is discussed, which is 
of importance for two-dimensional materials, and the associated beat patterns in squeezing are revealed.
\end{abstract}


\maketitle

\section{Introduction}
Background nuclear spins in a solid-state environment constitute a reservoir, which can manifest 
itself in two conflicting appearances. If left to its own devices, the spin bath becomes the primary decoherence 
channel of an embedded spin system \cite{khaetski02,merkulov02}, 
while under proper control it can turn into a resource for applications like quantum 
registers \cite{simon10,boehme12}, atomic-scale magnetometry \cite{zhao11,waldherr12,rondin14,wolf15}, 
or nanoscale imaging \cite{chekhovich12,mamin13,muller14,shi14}. Such regularization of nuclear spins 
has been enabled through advancements over the recent years in the field of quantum 
control \cite{vandersypen04,zhang-1407}. One particularly powerful tool is the dynamical 
decoupling of the system from decoherence channels using strong and fast pulse sequence 
protocols \cite{viola99,uhrig07,du09,biercuk09,lange10,vandersar12,ahmed13,casanova15}.
Another fruitful approach is to utilize dynamic nuclear polarization by means of
optical orientation of the electron spin and the enhanced hyperfine interaction 
through which nuclear spins can be polarized \cite{rudner07,urbaszek13}. As exemplary studies under this
general scheme, cooling of nuclear spins using Overhauser-field selective coherent population trapping 
\cite{issler10}, or suppression via the hole-assisted dynamic nuclear polarization feedback mechanism \cite{sun12} 
can be mentioned. Yet, another proposal to reduce nuclear spin fluctuations is to generate spin squeezing 
through unitary evolution in the presence of dynamic nuclear polarization \cite{rudner11}.

As a matter of fact in ultracold atoms, squeezing has become a well-established technique in controlling spin 
fluctuations \cite{ma11}.
The seminal paper by Kitagawa and Ueda introduced two specific means for squeezing, the so-called one-axis twisting
(OAT) and two-axis countertwisting (TAC) \cite{kitagawa}. 
For the OAT model, the dependence on the initial state and the effect of decay due to spontaneous emission have been 
studied \cite{jin09}, as well as the presence of an axial magnetic field \cite{law01,jin-pra-07}.
Even though TAC has superior squeezing characteristics 
\cite{kajtoch15}, in spinor Bose-Einstein condensates it has only been indirectly realized by transforming OAT 
using pulse sequences, hence in this respect these can be classified as {\em dynamical} 
TAC \cite{liu11,shen13,zhang14,muessel15}. Very recently it has been theoretically asserted that 
non-Hermitian OAT in the absence of decay can reach the squeezing 
limit of the Hermitian TAC model \cite{lee14,wu15}. 

It can be fair to state that from its inception the spin squeezing community has been lured by the applications 
on spinor condensates \cite{ma11}, to the extend that the system of nuclear spins is largely overlooked. 
The latter may appear as if simply a low-angular momentum special case. However, this particular system has 
its own originalities, such as the quadrupole interaction (QI), which is operational on nuclei with spin 
angular momenta larger than $\hbar/2$ \cite{cohen57,das58}. Moreover, unlike atomic systems, the squeezing 
nonlinearity is not a collisional many-body effect, but rather of single-particle origin, natively existing 
in the form of quadratic terms in the QI Hamiltonian. In terms of its technological prominence, one potential 
application may be in nanoscale magnetometry \cite{zhao11,waldherr12,rondin14,wolf15}. The strong link 
between the two is recently established by a breakthrough in single-shot readout that has been achieved over 
the nitrogen nuclear spin which is itself a quadrupolar nucleus of $I=1$ \cite{neumann10,buckley10}. 
In this connection, the ability to keep a certain spin component below the standard quantum limit 
for an extended duration of time would be certainly desirable. 
In this way sub-shot-noise sensitivities can be pursued via nuclear spin squeezing  \cite{taylor08}, much like 
those successful counterparts in photonic \cite{caves81}, and atomic systems \cite{ma11}.

In this article, our primary objective is to theoretically explore the nature of QI squeezing of a 
quadrupolar nuclear spin. The inspiration for this work is the very recent experimental demonstration
by Auccaise {\em et al.} of the squeezing in $^{133}$Cs nuclei of spin 7/2 \cite{auccaise15}. 
In their analysis, as a proof of principle they have considered a simple QI Hamiltonian 
neglecting the biaxiality ($\eta$) of the electric field gradient (EFG). In nanostructures such 
as semiconductor quantum dots, the biaxiality of the EFG is  quite pronounced \cite{bulutay12,bulutay14}. 
Its presence, as we shall show, offers the opportunity to combine both OAT and TAC models. Therefore, 
it can be termed as mixed-axis twisting (MAT) as was coined within a model context \cite{bian13}.
QI in its general biaxial form enables a crucial flexibility that can be harnessed for solid-state 
NMR-based quantum control purposes. 
Thus, in this work the behavior of QI squeezing is studied for various nuclear spin angular momenta and 
initial states, together with the corresponding squeezing speeds. 
Furthermore, additional effects of Zeeman interaction 
and dephasing on the steady-state squeezing are analyzed. Finally, the limiting case of 
extreme biaxiality ($\eta=1$) is separately treated, which can be realized in two-dimensional materials, 
and the interesting beat patterns in squeezing are identified.

Before we get into theoretical deliberations, some practical aspects pertaining to nuclear 
spin squeezing could be noteworthy. First of all, compared to atomic spin systems \cite{ma11}, 
the techniques for nuclear spins are markedly 
different on the experimental level, in particular owing to the pseudo-pure state framework of the NMR quantum 
processing \cite{gershenfeld97}. In spin squeezing studies, the standard starting point is the so-called
coherent spin state (CSS) that minimizes the uncertainties in the quadratures \cite{ma11}. With the 
current advancement of NMR techniques, creation of an initial CSS is no longer a problem, as has 
been demonstrated using strongly modulated pulse sequence, even in the presence of QI \cite{estrada13}. 
The same is true for the readout of the final state that is accomplished in NMR through the 
quantum state tomography \cite{chuang98}, which was also extended to quadrupolar nuclei \cite{teles07}.
Lastly, we would like to address the typical time scale to produce a sufficiently strong squeezing using QI. 
The key parameter in this context is the strength of the QI as characterized by the quadrupolar 
frequency. Typically this ranges from 10~kHz in lyotropic samples \cite{auccaise15} to 
a few MHz in self-assembled quantum dots \cite{bulutay12,bulutay14}. Based on these, one can estimate
the squeezing times to be on the order between 1-100~$\mu$s.

The paper is organized as follows. In Sec.~II we present the theoretical basis of our analysis
by providing the expressions for QI, spin squeezing measures, and different interaction Hamiltonians to be 
utilized later on. In Sec.~III we report the general trends of bare QI squeezing as a function of nuclear spin 
angular momenta, initial states, and EFG biaxiality; we also discuss how the squeezing rate is affected 
by these parameters. In Sec.~IV the steady-state QI squeezing is considered under dephasing and a static 
magnetic field. Sec.~V addresses specifically QI squeezing in the two-dimensional EFG case. Our main 
conclusions are summarized in Sec.~VI.

\section{Theory}
\subsection{Electric quadrupole interaction}
A nucleus with a spin angular momentum $I>1/2$ (in units of reduced Planck's constant, $\hbar$) possesses a
non-spherical charge distribution, hence has a non-zero electric quadrupole 
moment \cite{cohen57,das58}. This results in the coupling of the nuclear spin to the so-called electric 
field gradient (EFG), if available at that nuclear site. In a solid-state context, one common cause of EFG
is the crystal electric fields of polar group III-V semiconductor quantum dots under inhomogeneous 
strain~\cite{bulutay12,bulutay14}. The elements of the EFG tensor can be given by the Cartesian second 
derivatives of the (crystal) electric potential as,
$
V_{ij} \equiv \partial^2V/\partial x_i\partial  x_j \, .
$ 
Working in the frame of EFG principal axes, the convention is to label the coordinates such that 
$|V_{xx}|\leq |V_{yy}|\leq |V_{zz}|$, 
where $z$ ($x$) is referred to as the major (minor) principal axis of the EFG tensor~\cite{cohen57,das58}.

The QI of the nuclear spin $I$ with the EFG is described by the Hamiltonian \cite{cohen57}
\begin{equation}
\label{H-QI}
\hat{H}_Q = \frac{e^2qQ}{4I(2I-1)}\left[ 3\hat{I_z}^2 - \hat{I}^2 + \eta \frac{\hat{I}_+^2+\hat{I}_-^2}{2} \right]\, ,
\end{equation}
where, $e$ is the electronic charge,  $Q$ is the electric quadrupole moment, $\hat{I_z}$ is the $z$ component of
(dimensionless) spin angular momentum operator and $\hat{I}_{\pm}=\hat{I}_x\pm i\hat{I}_y$ are the standard spin 
raising/lowering operators. Two important parameters of the QI Hamiltonian are: $eq\equiv V_{zz}$, which is the major 
principal value of the EFG, and $\eta=\left(V_{xx}-V_{yy}\right)/V_{zz}$, which represents the asymmetry of the EFG, also known
as the EFG biaxiality parameter; the condition $\sum_i V_{ii}=0$, restricts  $\eta$ to the range between 
0 and 1 \cite{cohen57,das58}.

\subsection{Spin Squeezing}
Spin squeezing amounts to reducing quantum fluctuations below the standard quantum limit in one quadrature 
at the expense of the other quadrature so that overall Heisenberg uncertainty condition is not violated 
\cite{ma11}. In mathematical terms, for a spin vector 
having mutually orthogonal components $I_i$, $I_j$, $I_k$, 
if either $\langle \Delta \hat{I}_i^2 \rangle <\frac{1}{2} | \langle \hat{I}_k \rangle |$, or (exclusively) 
$\langle \Delta \hat{I}_j^2 \rangle <\frac{1}{2} | \langle \hat{I}_k \rangle | $ 
while respecting
$
\langle \Delta \hat{I}_i^2 \rangle \langle \Delta \hat{I}_j^2 \rangle \geq \frac{1}{4} | \langle \hat{I}_k \rangle |^2\, ,
$
then it corresponds to a spin-squeezed state \cite{ma11}. Here $\langle \Delta \hat{I}_i^2 \rangle$ corresponds to
the variance in $\hat{I}_i$.

To quantify the degree of squeezing, Kitagawa and Ueda~\cite{kitagawa} proposed the following squeezing parameter
\begin{equation}
\label{squeezing_parameter}
\xi_S = \frac{(\Delta I_n)_{min}}{\sqrt{I/2}},
\end{equation}
where, $(\Delta I_n)_{min}^2$ is the minimum variance of the spin component $\hat{I}_n$, 
with $\hat{n}$ being the unit vector perpendicular to mean spin direction $\langle \hat{\vec{I}} \,\rangle$. 
$\xi_S=1$ corresponds to CSS \cite{ma11}, and a value $\xi_S<1$ indicates a spin squeezed state. 
The minimum variance corresponds to
\begin{equation}
\label{min_variance}
(\Delta I_n)^2_{min} = \frac{C - \sqrt{A^2+B^2}}{2}\ ,
\end{equation}
where, $A=\langle \hat{I}_1^2 -\hat{I}_2^2 \rangle$, $B=\langle \hat{I}_1\hat{I}_2 + 
\hat{I}_2\hat{I}_1 \rangle$, $C=\langle \hat{I}_1^2 +\hat{I}_2^2 \rangle$ \cite{jin-pra-07,ma11}. Here, 
$\hat{I}_1=-\hat{I}_x\sin\phi+ \hat{I}_y\cos\phi$, 
and $\hat{I}_2=-\hat{I}_x\cos\theta\cos\phi-\hat{I}_y\cos\theta\sin\phi+\hat{I}_z\sin\theta$ are the two spin 
operators mutually perpendicular to mean spin orientation along the unit vector with Cartesian components
$(\sin\theta\cos\phi,\sin\theta\sin\phi,\cos\theta)$.

Even though we shall be using the above squeezing parameter $\xi_S$, we should mention that there are other estimates 
for squeezing \cite{ma11}, a popular variant being that introduced by Wineland {\em et al.} \cite{wineland}, 
related to $\xi_S$ as
\begin{equation}
\label{wineland}
\xi_R = \frac{I}{|\langle \hat{I} \rangle|} \xi_S.
\end{equation} 
$\xi_R$ is generally preferred in quantum metrology or in relating squeezing to entanglement \cite{sorensen01}. 
As these aspects are left out of the scope of this work, we shall be solely using the $\xi_S$ measure for squeezing.

Temporal change in squeezing can be monitored through the probability amplitudes, $|C_m|^2$ of each spin projection, 
$| I,m\rangle$, that can be simply extracted from the evolving state vector $| \psi(t)\rangle$ using \cite{jin-pra-07}
\begin{equation} \label{c_m_prob}
| \psi(t)\rangle=\sum^{I}_{m=-I} C_m(t) | I,m\rangle\, .
\end{equation}
Additionally, we shall be resorting to spin Wigner distribution to visualize the degree of squeezing along the two quadratures
\cite{qutip1,qutip2}.  

\subsection{Different squeezing Hamiltonians}
In the pioneering paper \cite{kitagawa}, two schemes of generating spin squeezing were introduced: 
the one-axis twisting (OAT) Hamiltonian

\begin{equation}
\label{H-OAT}
\hat{H}_{\hbox{\scriptsize OAT}} = \hbar\chi \hat{I}_z^2 \, ,
\end{equation}
and the two-axis countertwisting (TAC) Hamiltonian with one of its forms being
\begin{equation}
\label{H-TAC}
\hat{H}_{\hbox{\scriptsize TAC}} = \hbar\chi \left( \hat{I}_x^2 - \hat{I}_y^2 \right)\, ,
\end{equation}
where $\chi$ quantifies the squeezing amplitude.

It can be readily checked that the QI Hamiltonian, as given by Eq.~(\ref{H-QI}) in the presence of EFG biaxiality 
happens to be a combination of OAT and TAC Hamiltonians, which will be referred to as 
mixed-axis twisting (MAT) Hamiltonian \cite{bian13}
\begin{equation}\label{H-MAT}
\hat{H}_{\hbox{\scriptsize MAT}} = 
\frac{h f_Q}{6} \left[ 3\hat{I_z}^2 + \eta \frac{\hat{I}_+^2+\hat{I}_-^2}{2} \right]\, , 
\end{equation}
where in relation to the QI Hamiltonian, $h f_Q= 3e^2qQ/\left[ 2I(2I-1)\right]$, with $h$ being the Planck's constant,
and the $\hat{I}^2$ term has been dropped as it is constant for the spin $I$ nucleus under investigation.
The fact that for $\eta=0$, MAT degenerates to OAT case of Eq.~(\ref{H-OAT}) is quite obvious. At the other 
extreme, $\eta=1$,  it reduces to pure TAC, as it yields
\begin{equation}\label{H-eta1}
\hat{H}_{\hbox{\scriptsize MAT}}\left(\eta=1\right) = 
\frac{h f_Q}{6} \left[ \hat{I}^2+2\hat{I}_z^2-2\hat{I}_y^2 \right]\, ,
\end{equation}
which has the same form of Eq.~(\ref{H-TAC}), but around two countertwisting axes set by the two,
now degenerate, major principal directions, here $y$ and $z$. 
In other words, as illustrated in Fig.~1, when $\eta$ is swept from 0 to 1, the functionality of the QI 
Hamiltonian continuously transforms from OAT to TAC, as the EFG tensor changes from uniaxial 
(in this case, along the $z$ axis) to extreme biaxial character (along $z$ and $y$ axes).

\begin{figure}
 \includegraphics[width=1.05\columnwidth]{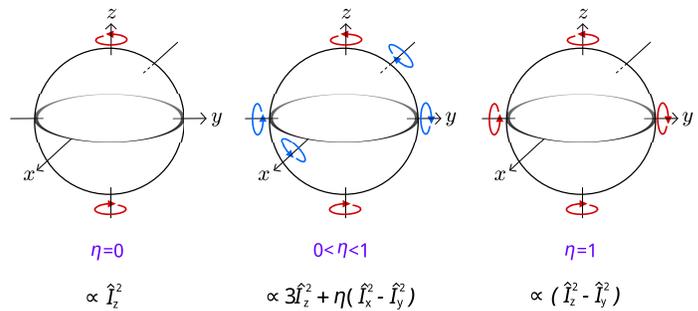}
\caption{(Color online) Successive transformation of QI Hamiltonian from OAT, over MAT, and to TAC as 
$\eta$ is swept within its interval, see Eqs.~(\ref{H-MAT}) and (\ref{H-eta1}). The twisting 
directions are indicated on relevant axes of the Bloch sphere.
}
\label{fig1}
\end{figure}

\subsection{Squeezing speed}
To quantify squeezing speed under various conditions we invoke two measures. As a general resort,
the Margolus-Levitin theorem sets a quantum speed limit based on the time it takes to evolve 
into an orthogonal state given by
\begin{equation}
\label{MLT}
\tau_\perp\ge\max\left\{ \frac{\pi\hbar}{2E}, \frac{\pi\hbar}{2\Delta E}\right\}\, ,
\end{equation}
where $E$ is the average energy as measured from its ground state level, and $\Delta E$ is the standard deviation 
of the system \cite{margolus98,giovanetti03}. 
In the general case of when the evolution is not toward an orthogonal state, a very similar expression 
is shown to be valid with the involvement of an extra prefactor that is based on the overlap between the initial 
and final states \cite{giovanetti03}.
As we shall be using CSS initial states, the variation with respect 
to $\eta$ of both $E$ and $\Delta E$ are very similar, and for the {\em initial} squeezing speed 
the average energy $E$ yields a good indicator.

A more direct measure is very recently provided by Opatrn\'{y} in the form of an explicit expression 
for the squeezing rate under a general twisting tensor \cite{opatrny15a}. In regard to our MAT case, 
it can be cast into
\begin{widetext}
\begin{equation}\label{squeeze-rate}
\mathcal{Q} =2I\sqrt{\left[\eta\cos2\varphi (1+\cos^2\vartheta)+3\sin^2\vartheta\right]^2+
4\eta^2\cos^2\vartheta\sin^2 2\varphi}\, ,
\end{equation}
\end{widetext}
where the spherical angles $(\vartheta,\, \varphi)$ define the initial CSS, which we
shall denote as $(\theta_{CSS},\, \phi_{CSS})$ in the remainder of the paper. 

\subsection{Zeeman and dephasing terms}
We shall additionally consider the effect of an external magnetic field having an arbitrary orientation in the EFG principal 
axes described by the spherical angles $\theta$ and $\phi$ as
\begin{equation} \label{H-zeeman}
\hat{H} = -\hbar \omega_0 \left( \hat{I}_x \sin\theta \cos\phi + \hat{I}_y \sin\theta \sin\phi + \hat{I}_z \cos\theta \right)\, ,
\end{equation}
where $\omega_0$ is the Larmor angular frequency. 
The inclusion of this Zeeman term to the aforementioned squeezing Hamiltonians turns them into the 
so-called Lipkin-Meshkov-Glick model \cite{lipkin65}, as was already employed in two-mode Bose-Einstein 
condensates \cite{vidal04,zhang14}. However, this time it refers to nuclei, which happens to be the main 
framework of the original model \cite{lipkin65}.

The dephasing of a nuclear spin will be accounted in our work within the well-known Lindblad formalism through 
the phase-flip channel model \cite{nielsen-book}
\begin{equation}
\begin{split} 
\frac{d}{dt} \hat{\rho}_S(t)  = &-\frac{i}{\hbar}\left[ \hat{H}, \hat{\rho}_S(t) \right] \\
& + W_{\phi} \left[\hat{I}_z\hat{\rho}_S(t)\hat{I}_z -\frac{1}{2}\left\{\hat{I}_z^2,\hat{\rho}_S(t)\right\}\right] \, ,
\end{split}
\end{equation}
where $\hat{\rho}_S$ is the spin system density operator, $W_{\phi}$ is the dephasing rate, $[\, ,]$ and $\{\, ,\}$ 
represent commutator and anti-commutator, respectively.

\subsection{The $\eta=1$ case}
The upper limit for $\eta$ is 1. So, one naturally wonders whether such a severe EFG biaxiality is actually practical. 
From the definition of $\eta$ and the condition $\sum_i V_{ii}=0$, it can easily be inferred that $\eta=1$ implies 
$V_{xx}=0$ and $V_{yy}=-V_{zz}$. 
The most common cause for EFG is the atomistic strain, described by a tensor $\epsilon_{ij}$.
If we assume the off-diagonal entries of $\epsilon_{ij}$ to be negligible in the EFG principal axes frame, then
the relation between EFG and strain diagonal entries is given by
\begin{equation}
V_{xx}=S_{11}\left[ \epsilon_{xx}-\frac{1}{2}\left( \epsilon_{yy}+\epsilon_{zz}\right)\right]\, ,
\end{equation}
and its successive cyclic permutations for $V_{yy}$ and $V_{zz}$. Here, $S_{11}$  in Voigt notation is a 
gradient elastic tensor component \cite{bulutay12}.

The case $V_{xx}=0$ can be attained non-trivially only for $\epsilon_{yy}=-\epsilon_{zz}$ together with $\epsilon_{xx}= 0$,
which also leads to $V_{yy}=-V_{zz}$, thus resulting in $\eta=1$. This strain combination directly suggests two-dimensional 
materials, which recently started to attract considerable attention \cite{wang12}. Within the considered EFG frame the material 
would lie along the $yz$ plane. Hence, a perpendicular static magnetic field needs to be $x$ oriented. In a following section we 
shall be considering this particular combination.


\section{Bare QI squeezing}
\subsection{General trends}
We first begin with the general squeezing trends of the bare QI Hamiltonian for quadrupolar nuclei with 
$I$ ranging from 1 to 9/2 \cite{note1}. 
As the initial spin state we start from CSS that is described on the Bloch sphere by the polar angle
$\theta_{CSS} = \pi /2$, and the azimuthal angle $\phi_{CSS} = \pi /2$. We follow the time evolution of 
such a single spin under OAT ($\eta =0$), MAT ($\eta =0.5$), and  TAC ($\eta =1$) Hamiltonians. 
As shown in Fig.~2, $I$=1 case is distinctly different from the others where perfect squeezing, $\xi_S =0$,
can be attained periodically at discrete instants for all $\eta$ values; this may be of importance 
for the nitrogen spins of NV centers. Also among the three models, 
TAC has the longest period. For $I>3/2$ we rather observe a quasi-periodic character under MAT and 
TAC for this initial state, whereas OAT always retains its periodicity with a linear frequency of $f_Q$
(see, Eq.~(\ref{H-MAT})).
In general, as $I$ increases the oscillation in $\xi_S$  increases, i.e., the (quasi-) period decreases. 

\begin{figure}
 \includegraphics[width=1.05\columnwidth]{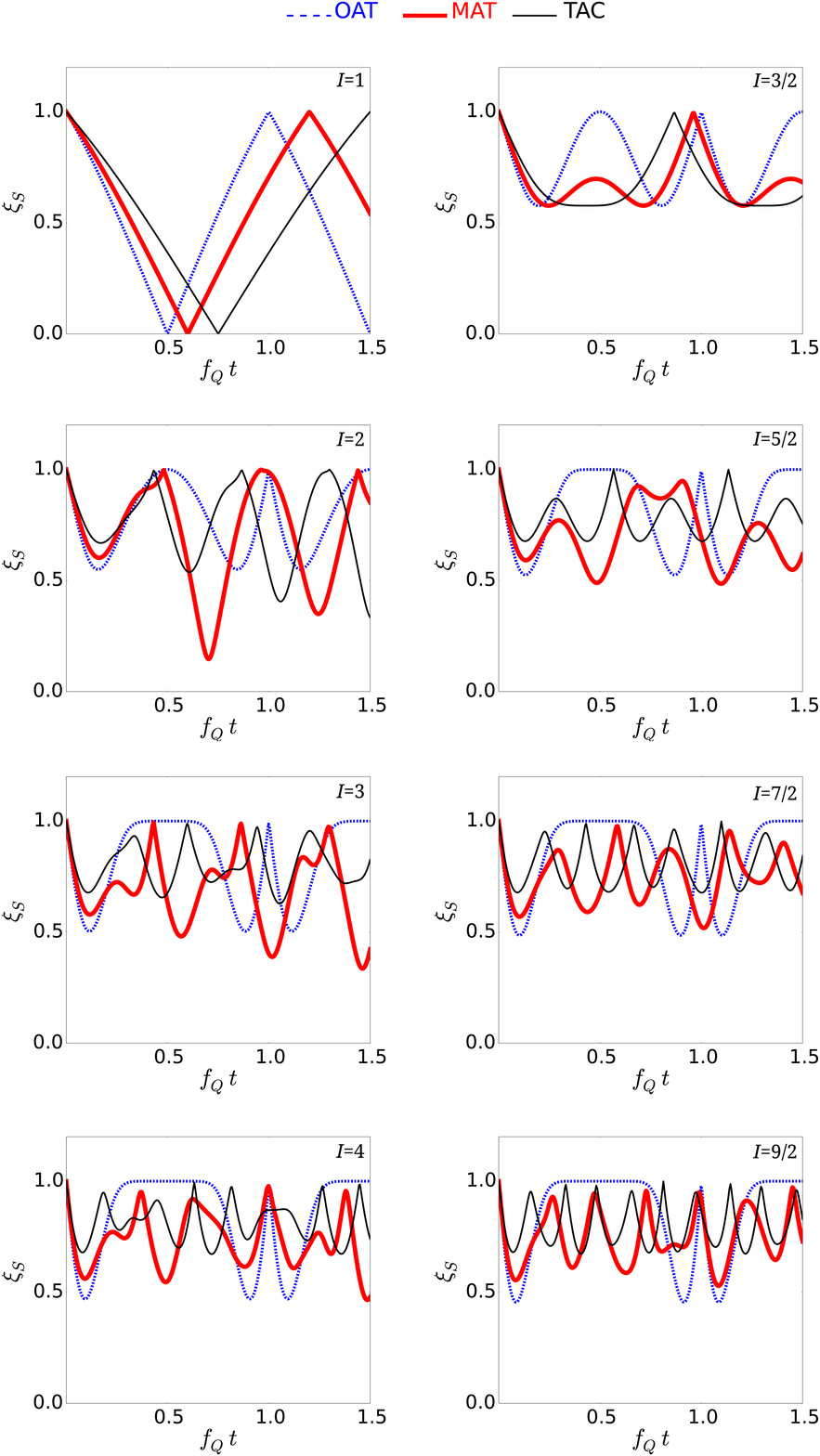}
\caption{(Color online) Comparison of the temporal squeezing characteristics of OAT, MAT  ($\eta =0.5$) 
and TAC models for quadrupolar nuclei with spin angular momenta from $I=1$ to 9/2, 
for $\theta_{CSS} = \pi /2$, and $\phi_{CSS} = \pi /2$. 
$f_Q$ represents the linear QI frequency, see Eq.~(\ref{H-MAT}).
}
\label{fig2}
\end{figure}

\begin{figure}
 \includegraphics[width=1.05\columnwidth]{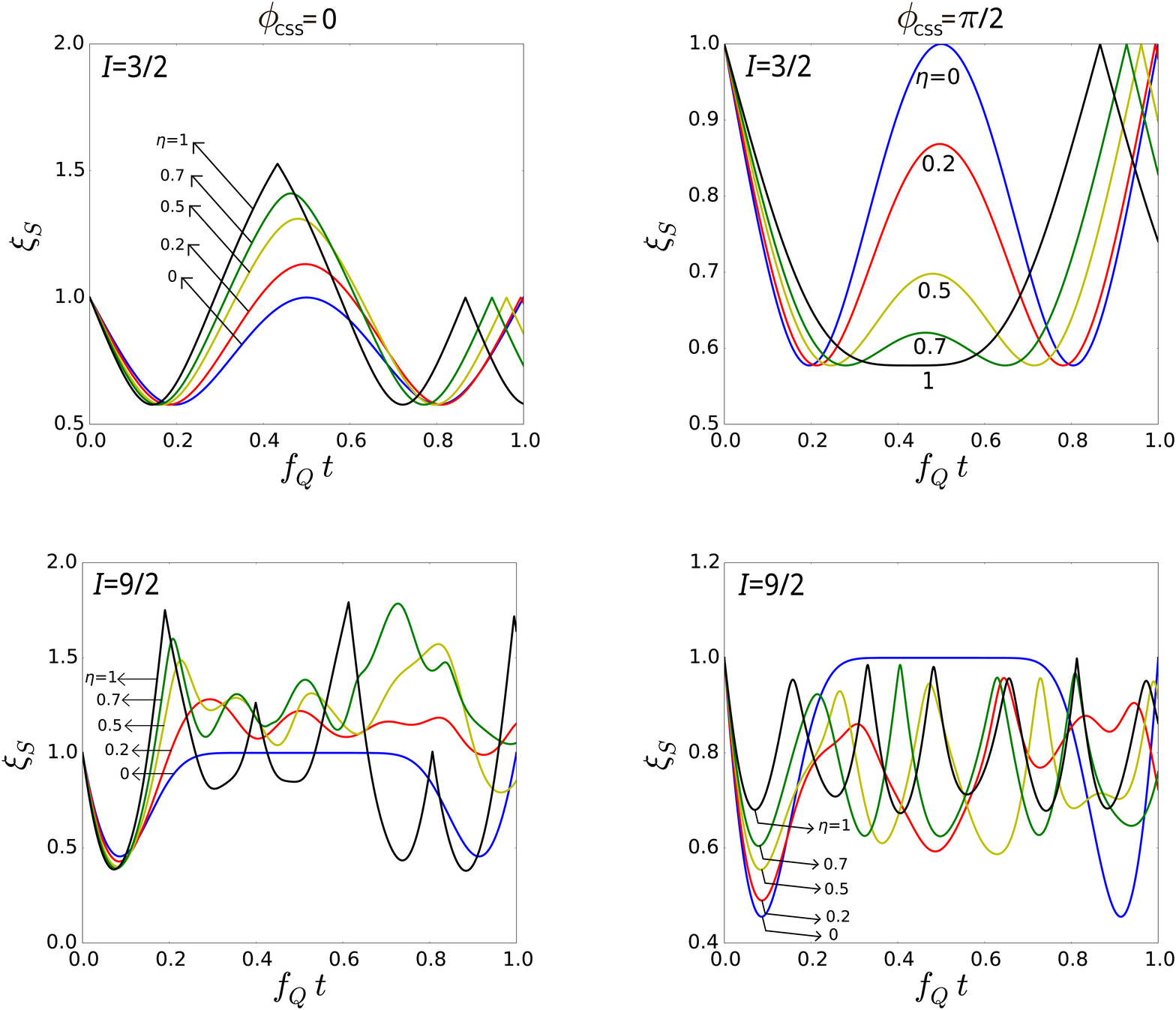}
\caption{(Color online) The EFG biaxiality parameter, $\eta$ dependence of spin squeezing 
for $I=3/2$ and 9/2. As the initial CSS, $\theta_{CSS}=\pi/2$ is used in all cases, and the left 
(right) column is for $\phi_{CSS}$ equal to 0 ($\pi/2$).
}
\label{fig3}
\end{figure}

In Fig.~3 the roles of EFG biaxiality, $\eta$  and the azimuthal angle of initial CSS, $\phi_{CSS}$ 
on the degree and rate of spin squeezing are demonstrated over $I=3/2$ and 9/2 spins. Starting with 
the former, keeping the polar angle of initial CSS fixed at $\theta_{CSS} = \pi /2$, and for 
$\phi_{CSS}=0$ (i.e., left panel of Fig.~3), as $\eta$ increases, squeezing more rapidly changes 
and even becomes anti-squeezed ($\xi_S >1$) in certain intervals, unlike the OAT case. This pattern 
is reversed for $\phi_{CSS}=\pi/2$ (i.e., right panel of Fig.~3) where  initial squeezing rate decreases 
with $\eta$, however bears the benefit that the spin in a particular quadrature stays squeezed for 
longer duration compared to OAT. The same also applies for $I=9/2$, and they are further accompanied with a 
change in the minimum $\xi_S$ value as a function of $\eta$. These observations hint that the optimal 
initial condition depends on the specific objective, such as fast initial squeezing (left panel), or 
prolonged squeezing (right panel). Further analysis will be presented in the following sections.

\begin{figure}
\includegraphics[width=1.05\columnwidth]{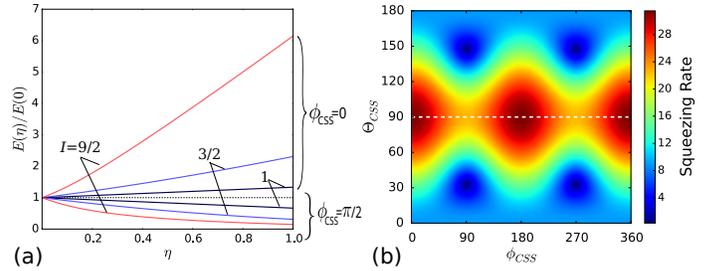}
\caption{(Color online) (a) Variation of CSS energy as a function of $\eta$ for various $I$, and two 
different $\phi_{CSS}$ values as indicated. $\theta_{CSS}=\pi/2$ is used in all cases, and 
energies are measured from their respective ground state energies (see, Eq.~(\ref{MLT})).
(b) Squeezing rate, $\mathcal{Q}$ variation over the full Bloch sphere for 
the representative case of $I=9/2$, and $\eta=0.5$ . White dashed line marks the equator.
}
\label{fig4}
\end{figure}

\subsection{Squeezing speed}
The reversal in the squeezing speed from $\phi_{CSS}=0$ to $\pi/2$ as identified above can be explained 
by the Margolus-Levitin theorem, stated in Eq.~(\ref{MLT}).
Since these initial states $(\theta_{CSS}, \phi_{CSS})$: $(\pi/2,0)$ and $(\pi/2,\pi/2)$, only 
differ by their positions on the equatorial plane of the Bloch sphere, so when the Hamiltonian 
becomes uniaxial, that is $\eta=0$, their average energies become identical. As a matter of 
fact as shown in Fig.~4(a), regardless of the spin $I$ value, for $\phi_{CSS}=0$  the variation 
of the initial CSS energy increases with $\eta$, while this is just the opposite for 
$\phi_{CSS}=\pi/2$, corroborating the associated squeezing speeds. Also, the energetic variation 
can be observed to be increasing with $I$. This can also be seen from the squeezing rate, $\mathcal{Q}$ 
expression in Eq.~(\ref{squeeze-rate}) which is directly proportional to $I$. That is to say, 
higher spin nuclei can potentially benefit from much faster squeezing.

To extend our discussion over the full Bloch sphere, we display in Fig.~4(b) the variation of the squeezing rate, 
given by Eq.~(\ref{squeeze-rate}), under the MAT Hamiltonian for $\eta=0.5$, and $I=9/2$. The squeezing rate minima 
(i.e., blind spots) occur at four locations over the $y-z$ plane (i.e., $\phi_{CSS}=\{\pi/2,\, 3\pi/2$\}):
for the $\eta=0.5$ example here, they lie very close to points  $\theta_{CSS}=\{\pi/6,\, 5\pi/6$\}.
The maximum squeezing rate takes place along the $\pm x$ direction which corresponds to the minor principal 
axis of the EFG tensor. Overall, it can be inferred from the same figure 
that OAT sets the baseline squeezing rate, and strong divergence in either direction from that value 
occurs primarily around the equator band.

\begin{figure}
 \includegraphics[width=1.05\columnwidth]{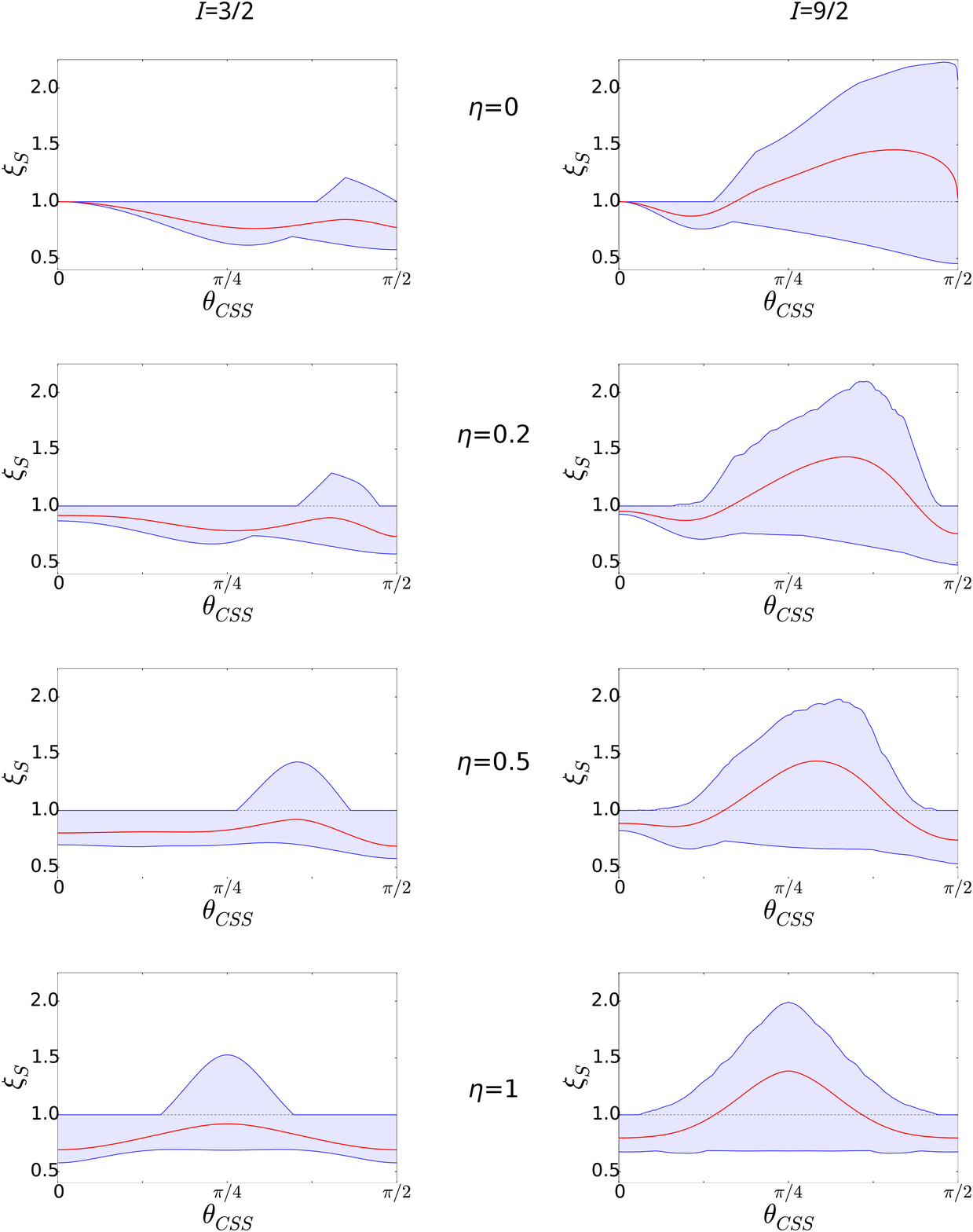}
\caption{(Color online) The interval over which the temporal squeezing changes as function of $\theta_{CSS}$
for $I=3/2$ and 9/2. $\phi_{CSS}=\pi/2$ is used in all cases. The (red) lines within the shaded zones 
represent the mean squeezing values.
}
\label{fig5}
\end{figure}	

\subsection{Dependence on the polar angle of the initial state}
Next, we explore the effect of the initial CSS polar angle, $\theta_{CSS}$ on the squeezing 
characteristics for distinct $\eta$ values. As shown in Fig.~5, OAT at $\theta_{CSS} =0$ (or $\pi$) yields no 
squeezing (a blind spot), but preserves its CSS character, a fact that is already known \cite{jin09,opatrny15b}. 
For all $\eta$ values anti-squeezing exists at specific $\theta_{CSS}$ intervals. Moreover, as $\eta$ increases the
interval retreats from the equatorial plane, reaching $\theta_{CSS}=\pi/4$ for $\eta=1$, for both $I=3/2$ and 9/2. 
Another observation is that the minimum accessible $\xi_S$ value for OAT is strongly dependent on the initial 
$\theta_{CSS}$, which becomes to a large extent independent of it as $\eta$ approaches unity. In other words, 
MAT evens out over the Bloch sphere the minimum level of $\xi_S$ for high $\eta$ values.

The mean squeezing values are also indicated  in Fig.~5 as red lines. Another way to look at this is the 
squeezed duty cycle, defined as $t_S/T$, where $t_S$ is the amount of time spin stays in the $\xi_s\le 1$ regime 
over a sufficiently long time span of $T$  \cite{huang12}. As shown in Fig.~6, the duty cycle decreases as either 
$I$ or $\eta$ increases, confirming Fig.~5. 
For $I=9/2$ and $\eta=1$ case, efficient squeezing zone gets confined to either $\theta_{CSS}=0$ or $\pi/2$ neighborhoods. 
These zones widen for the $I=3/2$ case. That means, it is harder to achieve large squeezing duty cycles for higher 
spin nuclei under wide initial CSS conditions.

\begin{figure}
 \includegraphics[width=1.05\columnwidth]{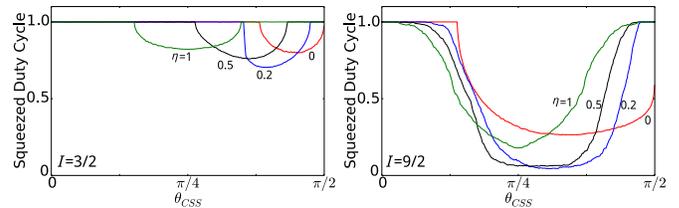}
\caption{(Color online) The variation of squeezed duty cycle with respect to $\theta_{CSS}$
for $I=3/2$ and 9/2, and various $\eta$ values. $\phi_{CSS}=\pi/2$ is used in all cases. 
}
\label{fig6}
\end{figure}

\begin{figure}
 \includegraphics[width=1.05\columnwidth]{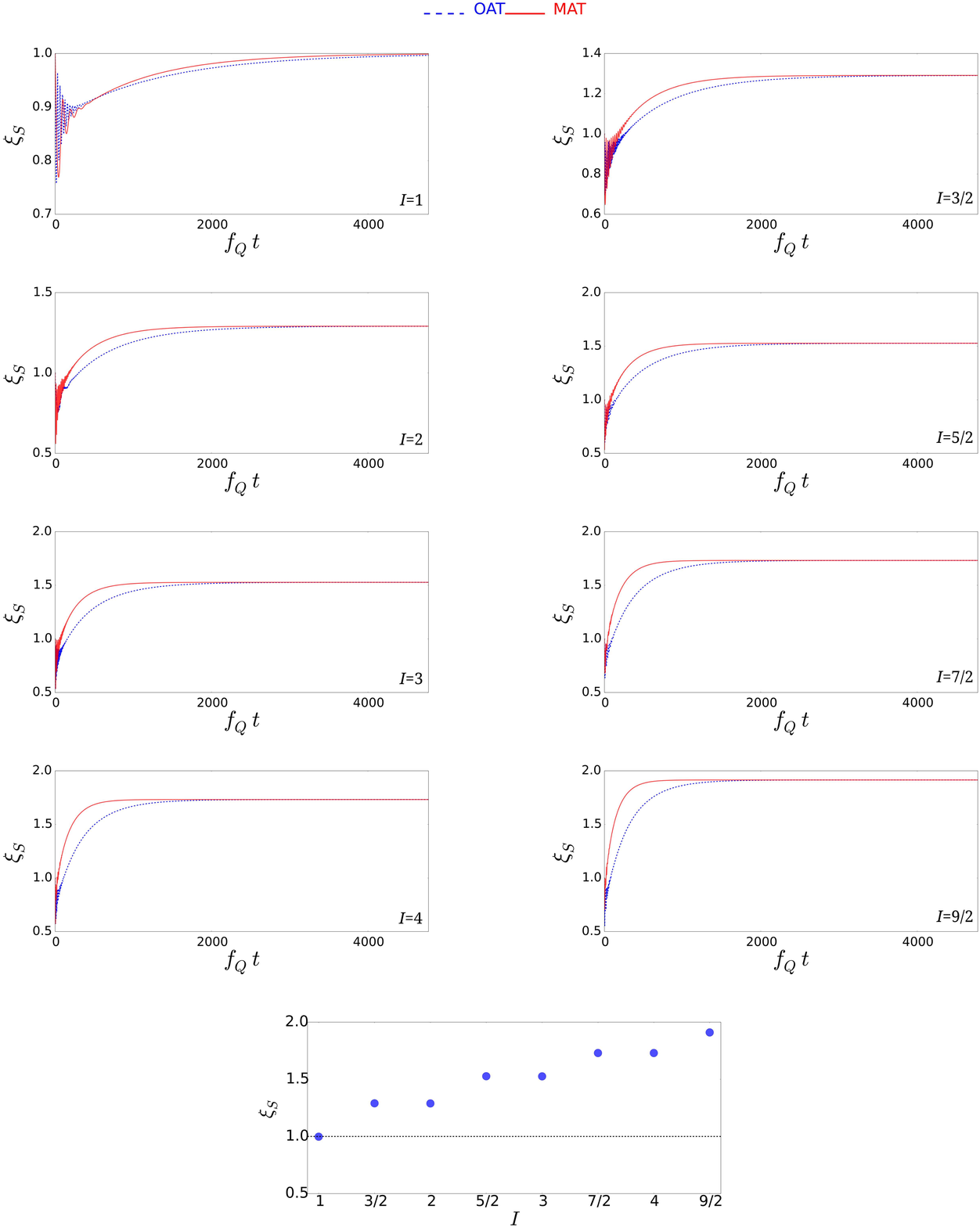}
\caption{(Color online) The approach to steady-state due to dephasing for OAT and MAT ($\eta=0.5$) models under an $x$-directed
static magnetic field. See text for the values. The bottom plot shows the corresponding steady-state values for various $I$.
}
\label{fig7}
\end{figure}	

\section{Effect of Zeeman interaction and dephasing on steady state squeezing}
Having considered the squeezing characteristics of the bare QI Hamiltonian, we now include the static magnetic 
field and the dephasing terms. We are particularly interested in the progression of squeezing toward the steady 
state under these general conditions. The magnetic field is chosen to be along $x$ direction with an
associated linear Larmor frequency equal to that of QI, $\omega_0/2\pi= f_Q$. The dephasing rate is taken as $W_\phi/2\pi=0.001 f_Q$. 
However, we should mention that these particular choices are not that critical for steady state characteristics. 
In Fig.~7 (upper panels) we compare the behaviors of OAT and MAT ($\eta=0.5$) for $I=1$ to 9/2. We observe that for
a given $I$ both OAT and MAT reach to the same steady state squeezing. In dynamics, MAT shows more oscillations,
and as $I$ increases it attains the steady state value faster than OAT. Their distinction are somewhat reminiscent of 
the underdamped and overdamped responses for MAT and OAT, respectively. In the bottom panel of Fig.~7 the
steady-state values are plotted, which exhibits a step-wise increase in anti-squeezing with respect to $I$.
Only for $I=1$ case, $\xi_S=1$ is realized, and for higher spins terminal states get anti-squeezed, the degree of which 
increases with $I$. This points out the adverse effect of dephasing in keeping the quadrupolar spin in a squeezed state.
On the other hand, recent studies on non-Hermitian TAC \cite{lee14}, and OAT \cite{wu15} models have reported 
rather favorable squeezing conditions with respect to their Hermitian counterparts. The discrepancy may be caused by the fact that 
their models involved a dissipative channel, whereas we have a nondissipative phase-damping decoherence \cite{ma11}.
\begin{figure}
 \includegraphics[width=1.05\columnwidth]{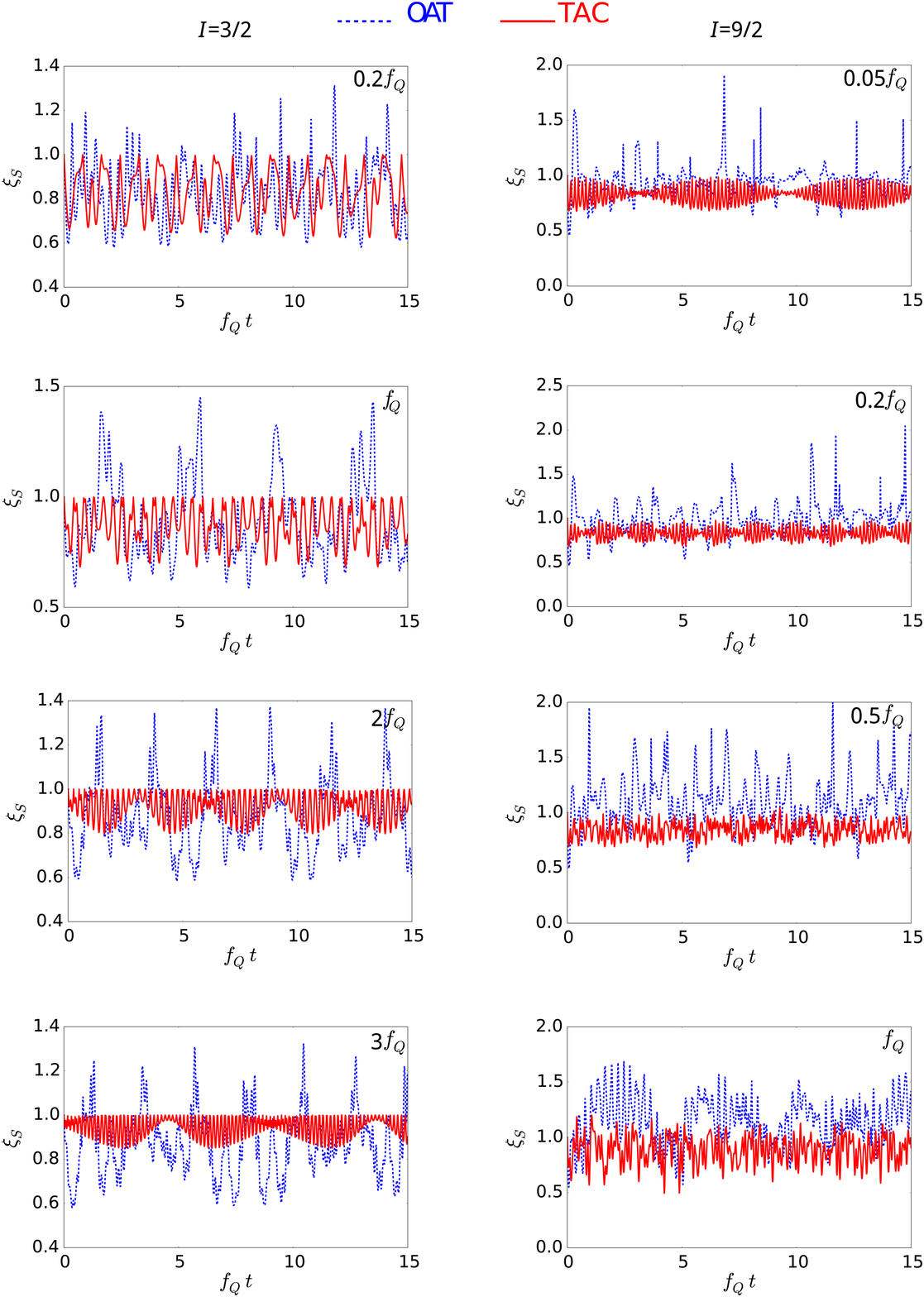}
\caption{(Color online)  The comparison of the temporal squeezing patterns for OAT, and the $\eta=1$ TAC models 
under an $x$-directed static magnetic field for various linear Larmor frequencies, $\omega_0/2\pi$ as labeled in each panel.
 $f_Q$ represents the linear QI frequency, see Eq.~(\ref{H-MAT}).
}
\label{fig8}
\end{figure}

\begin{figure*}
 \includegraphics[width=1.5\columnwidth]{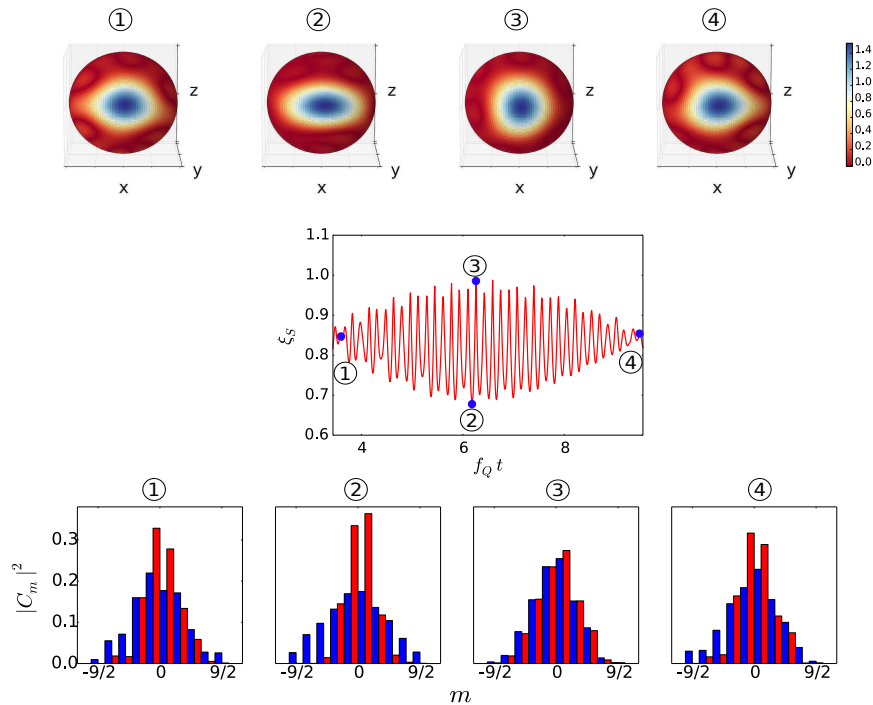}
\caption{(Color online)  Spin Wigner distributions (top row) at four different time instants over the beat 
pattern as numbered (center row), and the corresponding probability amplitudes (bottom row) for the
squeezed (light/red) and anti-squeezed (dark/blue) quadratures. An $I=9/2$ spin for $\eta=1$, TAC model 
under an $x$-directed static magnetic field having a linear Larmor frequency of $0.05\, f_Q$ is considered.
}
\label{fig9}
\end{figure*}
\section{$\eta=1$ case and the beat pattern}
As we  have discussed in the Theory section, $\eta=1$ case corresponds to the practically important case of two-dimensional
materials under inhomogeneous in-plane strain. So, now we consider the case for  $\eta=1$  in conjunction with a 
magnetic field along $x$ direction, which would be perpendicular to the two-dimensional material plane having 
$V_{yy}$ and $V_{zz}$ EFG components. As the initial spin state we again choose a  CSS with 
$\theta_{CSS} = \phi_{CSS} = \pi /2$. The squeezing patterns for $I=3/2$ and 9/2 shown in Fig.~8 are quite distinct 
from the previous cases. Namely, a beat pattern in squeezing appears for TAC with $\eta=1$, which is not the 
case for OAT. The latter quite frequently becomes anti-squeezed, whereas  $\eta=1$  QI Hamiltonian confines the
nuclear spin largely in the squeezed regime. The beating in TAC arises due to two tones originating from the Larmor 
precession under the external magnetic field and the biaxial QI term. For the higher spin ($I=9/2$) it becomes manifest 
at a lower Larmor frequency. Another intriguing feature is that the $I=3/2$ case (at variance to $I=9/2$) displays a 
squeezed beat pattern, which is {\em clipped} from above at $\xi_S=1$.
We should note that for {\em integer} spins, a beat pattern can also arise for TAC in the absence of an external magnetic 
field when an initial CSS is chosen at the polar region of the Bloch sphere.

To improve our understanding, we examine a portion of the time series from the $I=9/2$ TAC beat pattern 
in Fig.~8 at the linear Larmor frequency of $0.05\, f_Q$, separately shown on the center row of Fig.~9.
First, we compare the probability amplitudes, $\left | C_m\right |^2$ of each spin projection 
$|I,m\rangle$ for $I=9/2$ (see, Eq.~(\ref{c_m_prob})) at four time instances over the beat pattern, 
as marked on the center panel. It can be observed that $\xi_S$ values correlate well with the variation 
of the squeezed quadrature, shown as light/red bars in Fig.~9. The three-dimensional Wigner 
function plots on the top row provide a further insight for these cases. Here, the maximum beatings 
that occur at the first and fourth instants, are reflected by their rather conspicuous Wigner 
distributions. Those for the second and third instants are quite different 
from these, which represent the expected maximum squeezing and CSS behaviors, respectively.

\section{Conclusions}
In this work, quadrupolar nuclear spin squeezing is studied through its native general QI Hamiltonian,
which for a non-zero $\eta$ value corresponds to MAT model. Its main tenet is that as the EFG tensor 
changes from uniaxial to extreme biaxial character, the functionality of the QI Hamiltonian 
continuously transforms from OAT, over MAT, to the TAC squeezing models.
Compared to OAT, as $\eta$ increases, MAT  evens out the minimum level of $\xi_S$ over the Bloch sphere.
In regard to initial states, we reported the preferable cases depending on the specific aim
for either the speed, or the duration of the squeezing.
By including a phase-damping channel, the steady-state characteristics are also investigated 
which indicates that terminal states get exceedingly anti-squeezed as $I$ increases. This 
exemplifies an adverse effect of dephasing in retaining a quadrupolar spin in the squeezed regime.
As a matter of fact, even in the absence of dephasing, achieving large squeezing duty cycles becomes 
harder for higher $I$ values, being restricted to initial CSS around polar or equatorial bands. 
Finally, for two-dimensional materials possessing $\eta\rightarrow 1$, and subject to perpendicular 
magnetic field, a beating in squeezing is predicted, which arises due to two tones originating 
from the Larmor precession under the external magnetic field and the biaxial QI term.
In general terms, we believe that this field deserves further attention due to its potential impact on 
magnetometry, two dimensional systems, or simply as a means for quantum control over the nuclear 
spins within a solid-state host matrix.

\begin{acknowledgments}
We are grateful to S. Turgut, M. \"{O}. Oktel, and D. Suter for their valuable comments.
This work was supported by T\"UB\.ITAK, The Scientific and Technological Research Council 
of Turkey through the Project No. 112T178.
The numerical calculations reported in this paper were partially performed at T\"UB\.ITAK ULAKB\.IM, 
High Performance and Grid Computing Center (TRUBA resources).
\end{acknowledgments}

\bibliography{manuscript}

\begin{thebibliography}{67}
\bibitem [{not()}]{note1}%
  \BibitemOpen
  \href@noop {} {}\bibinfo {note} {We should note that there are also stable
  nuclei with $I>9/2$, such as $^{50}$V of $I=6$. But these are of very low
  abundance, and the trend that comes out of our analysis also applies to such
  higher spins.}\BibitemShut {Stop}%
\end{thebibliography}

\end{document}